\documentclass[preprint,journal]{vgtc}            % preprint (journal style)

\vgtccategory{Research}

\vgtcpapertype{algorithm/technique}

\newif\ifshowcomments
\showcommentsfalse

\ifshowcomments

\newcommand{\s}[1]{{\color{red}\st{#1}}}

\newcommand{\rk}[1]{{\color{blue} [RK: #1]}}
\newcommand{\pp}[1]{{\color{orange} [PP: #1]}}
\else
\newcommand{\s}[1]{}

\newcommand{\pp}[1]{}
\newcommand{\rk}[1]{}
\fi

\title{Neighborhood-Preserving Voronoi Treemaps}

\author{%
    \authororcid{Patrick Paetzold$^*$}{0000-0002-1315-4602},
    \authororcid{Rebecca Kehlbeck$^*$}{0000-0002-0095-5865},
    \authororcid{Yumeng Xue}{0000-0002-8195-517X},
    Bin Chen,
    \authororcid{Yunhai Wang}{0000-0003-0059-6580},
    \authororcid{Oliver Deussen}{0000-0001-5803-2185}
}

\authorfooter{
    \item
  	$^*$ These authors contributed equally to this work.
   \item
  	P. Paetzold, R. Kehlbeck, Y. Xue, B. Chen, and O. Deussen are with University of Konstanz.
  	E-mail: \{firstname.lastname\}@uni-konstanz.de.
      \item
  	Y. Wang is with Renmin University of China, Beijing, China.
  	E-mail: wang.yh@ruc.edu.cn.

}

\abstract{%
 Voronoi treemaps are used to depict nodes and their hierarchical relationships simultaneously. However, in addition to the hierarchical structure, data attributes, such as co-occurring features or similarities, frequently exist. Examples include geographical attributes like shared borders between countries or contextualized semantic information such as embedding vectors derived from large language models. 
In this work, we introduce a Voronoi treemap algorithm that leverages data similarity to generate neighborhood-preserving treemaps. 
First, we extend the treemap layout pipeline to consider similarity during data preprocessing. 
We then use a Kuhn–Munkres matching of similarities to centroidal Voronoi tessellation (CVT) cells to create initial Voronoi diagrams with equal cell sizes for each level. 
Greedy swapping is used to improve the neighborhoods of cells to match the data's similarity further.
During optimization, cell areas are iteratively adjusted to their respective sizes while preserving the existing neighborhoods.
We demonstrate the practicality of our approach through multiple real-world examples drawn from infographics and linguistics. To quantitatively assess the resulting treemaps, we employ treemap metrics and measure neighborhood preservation.
}


\teaser{
  \includegraphics[width=\linewidth]{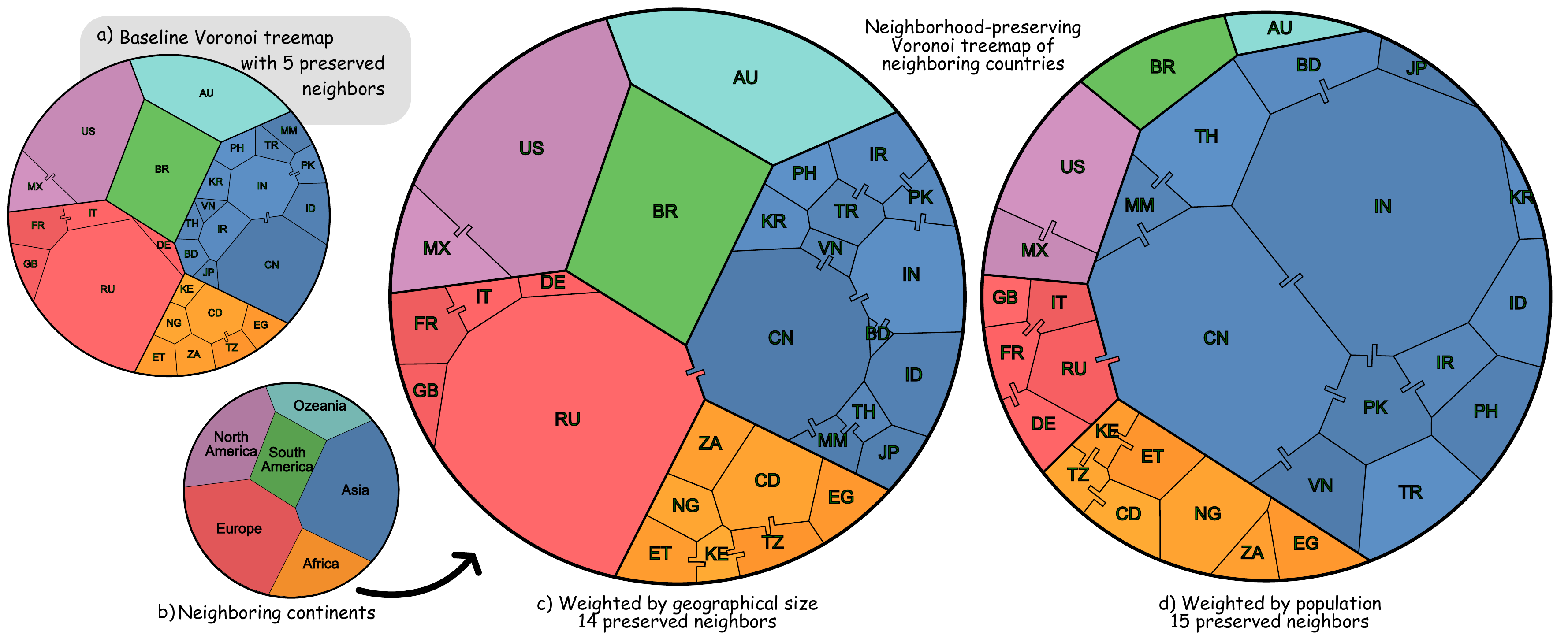}
   \centering
  \caption{%
    Our method creates Voronoi treemap layouts optimized to preserve similarities between nodes, resulting in appropriate neighborhoods in the treemap visualization.
   For this example, we use a two-level hierarchy dataset of large countries with a population of $>$ 50 Mio (+ Australia), grouped by their continents.
   First, we filter existing similarity between nodes, in this case, the geographic neighborhood between countries, to create similarity constraints.
   Next, Voronoi diagrams are initialized on each level using a combination of dimensionality reduction, matching, and cell swapping.
   During optimization, based on the constraints, an attraction force moves nodes towards other similar nodes, so their Voronoi cells share edges.
   The final Voronoi treemaps (c)-(d) show that most of the similarity constraints, representing geographical neighborhood, have been realized (see the \enquote{Interlocking} metaphor). Each continent is assigned a color, and countries of a continent have slightly varied colors from their neighbors.
   }
\label{fig:teaser}
}

\graphicspath{{figs/}{figures/}{pictures/}{images/}{./}} % where to search for the images

\usepackage{tabu}                      % only used for the table example
\usepackage{booktabs}                  % only used for the table example
\usepackage{lipsum}                    % used to generate placeholder text
\usepackage{mwe}                       % used to generate placeholder figures

\usepackage{mathptmx}                  % use matching math font

\usepackage{float}
\usepackage{csquotes}

\usepackage{microtype}                 % use micro-typography (slightly more compact, better to read)
\PassOptionsToPackage{warn}{textcomp}  % to address font issues with \textrightarrow
\usepackage{textcomp}                  % use better special symbols
\usepackage{mathptmx}                  % use matching math font
\usepackage{times}                     % we use Times as the main font
\usepackage[linesnumbered, ruled, noend]{algorithm2e}
\usepackage{amsmath}
\usepackage{wrapfig}
\usepackage{amssymb}

\usepackage[labelfont=sf,textfont=sf]{subcaption}
\usepackage[pass]{geometry}
\usepackage{multirow}
\usepackage{makecell}
\usepackage{ccicons}
\usepackage{makecell}
\usepackage{array}
\usepackage{multirow}
\usepackage[export]{adjustbox}% http://ctan.org/pkg/adjustbox
\usepackage{soul}

\begin{document}

%%%%%%%%%%%%%%%%%%%%%%%%%%%%%%%%%%%%%%%%%%%%%%%%%%%%%%%%%%%%%%%%
%%%%%%%%%%%%%%%%%%%%%% START OF THE PAPER %%%%%%%%%%%%%%%%%%%%%%
%%%%%%%%%%%%%%%%%%%%%%%%%%%%%%%%%%%%%%%%%%%%%%%%%%%%%%%%%%%%%%%%

\firstsection{Introduction}

\maketitle

Treemaps are a popular visualization method for hierarchical structures wherein each child element is fully enclosed within the borders of its parent element, adhering to what is commonly referred to as the containment property~\cite{treemap_taxonomy}. 
Balzer and Deussen~\cite{origVoronoiTreemap} introduce the Voronoi treemap approach, which subdivides non-rectangular regions into polygons. Voronoi Treemaps are widely used 
to underscore the inherent hierarchical structure within the dataset. Such hierarchical datasets are prevalent across various sectors, like economic datasets with stock market categories, file structures, or geographic data. 

However, in addition to their hierarchical arrangement, nodes may have relations to other nodes with the same parent, across parents, or even across multiple levels.
Examples are countries that share a border, files of the same type or similar genres across movies.
From other well-established visualization research, such as Gestalt laws~\cite{wertheimer1938laws}, it is known that using the proximity of data items to express their similarity is beneficial. This was our motivation to utilize the similarity of nodes during the optimization of Voronoi treemaps.
Given a real-valued similarity value between two nodes, we want to encode this relation on the distance between the two cells. In an ideal case, this means that for very similar nodes, their cells become \textit{neighbors}, sharing a cell border. We aim to \textit{preserve} this shared cell border during the optimization. 
 
To maximize the impact of similarities on the 2D layout, we reformulate the steps of the traditional Voronoi treemap pipeline~\cite{Scheibel2020TaxonomyTreemapVisualization}, integrate virtual nodes, propose operations such as swapping, and an updated optimization step that utilizes data similarity to move cells and preserve the neighborhood of similar cells.

Although some previous works adapted parts of the pipeline, to our knowledge, no method exists that integrates similarities in all pipeline steps without any given external positions. 
Buchin et al.~\cite{similarity_treemaps} propose a method to preserve neighborhoods in 2-level rectangular treemaps but assume that global and/or local cell positions and neighborhoods have been created using another algorithm.
Nocaj and Brandes~\cite{Reference_Map} proposed a Voronoi treemap layout that uses the similarity of hierarchy nodes to create a similarity graph. However, their method depends on a heuristic, dataset-specific parameter to reduce the edges in the graph. The layout of the so-called Reference Voronoi treemap~\cite{Reference_Map} is then initialized using a combination of MDS and scaling to the parent cell. 
They do not consider similarity during the optimization step.
Similarly, Yao et al.~\cite{PowerHierarchy} create Voronoi treemaps of dynamic hierarchies, where the hierarchical structure may change between time steps. The initial diagram is created using external, given positions for each cell. Each step is initialized with the scaled positions of the previous time step. Nodes whose parents changed are reinserted within the new parent to initialize each time step via projection. However, each node is randomly assigned to a Voronoi cell, and data similarity is not considered.
To summarize, none of the previous work utilizes similarity during the final optimization steps, where cells move. However, as shown by Nocaj et al.~\cite{Reference_Map}, up to 50\% of the geometric neighborhoods may change again.
In contrast, our method integrates and leverages similarity in all steps of the generation pipeline.
Similarities between nodes of each hierarchy level can be represented via an adjacency matrix. Another view on this data is as a graph, where nodes are linked if their similarity exceeds zero. This may result in a fully connected graph.
The treemap's dual graph must be a planar graph~\cite{similarity_treemaps}. Creating such a planar layout from a fully-connected graph is not trivial; therefore, we do not layout this graph directly, but instead utilize binning and filtering of the node similarities to reduce the complexity.

Our pipeline (\autoref{fig:exteded_treemap_pipeline}, \autoref{fig:treemap_pipeline_level}) consists of multiple steps: (1) The complexity of the node similarities is reduced by binning and filtering, extracting \textit{constraints} that represent the most critical similarities between the nodes. (2) The 2D position of each node is calculated using a dimensionality reduction technique and matched to a centroidal Voronoi tessellation node (CVT) whose distances to other cells best match the nodes' \textit{similarities}. 
After creating the Voronoi cells, we further improve cell \textit{neighborhoods} by \textit{swapping cells}.
(4) During the subsequent Voronoi optimization, all Voronoi diagrams on a hierarchy level move cells to maximize the distances between cell centroids. 

We adapt the relaxation step with an additional \textit{similarity} force that moves cells towards other similar cells, so that they become \textit{neighbors} in the Voronoi treemap and share an edge.
In addition, cells may grow in size or shrink so that the cell area represents a data attribute.

\noindent In summary, the main contributions of this paper are:

\begin{itemize}
    \setlength\itemsep{0.25em}
    \item A new hierarchy pre-processing step and Voronoi layout initialization using similarity-based matching and swapping
    \item An adapted Voronoi optimization method with similarity-based attraction forces during optimization to preserve Voronoi cell neighborhoods
     \item An \enquote{Interlocking} metaphor on Voronoi treemap edges as a direct encoding of similarity as neighborhood
     \item A qualitative and quantitative evaluation of application datasets using synthetic and real-world datasets across several domains
\end{itemize}

\begin{table*}[t]
    \centering
    \resizebox{\linewidth}{!}{
    \begin{tabular}{|c|c|c|c|c|c|c|c|c|c|}
     \hline
        Algorithm 
        &  Initial Layout
        & \thead{ Similarity Data  Attribute} 
        & \thead{ Adj.-based Geom.  Init}
        & \thead{ Adj. Geom.  Optim.} 
        & \thead{ Space  filling}
        & \thead{ Area  Proportional}
        & Levels 
        & \thead{ Cell  Shape} \\
     \hline
          Rec Map \cite{RecMap} & external & no& yes& -& yes& yes & 1 & rectangle\\
          \hline
        Adj.-Preserv. Treemaps\cite{similarity_treemaps} & external& no& yes& -& no& no & 2 & rectangle\\
    \hline
          Iso Match \cite{IsoMatch}& computed& yes& -& -& yes& no & 1 & rectangle\\
                       \hline
          Gosper Map \cite{GosperMap} & -& no& yes& -& yes&yes & $\geq$ 1 & polygon\\
          \hline
                  Reference Maps \cite{Reference_Map} & computed& yes& yes& no& yes& yes & $\geq$ 1 & polygon\\
         \hline
         Power Hierarchy\cite{PowerHierarchy} & computed& no& no& no& yes& yes  & $\geq$ 1 & polygon\\
         \hline
         Nmap\cite{Nmap} & computed& yes& yes& no& yes& yes& $\geq$ 1 & rectangle\\
         \hline
         Weighted Maps\cite{ghoniemWeightedMapsTreemap2015} & external& yes& yes& no& yes& yes& $\geq$ 1 & rectangle\\
         \hline
        CodeSurveyor\cite{hawesCodeSurveyorMappingLargescale2015} & computed& yes& yes& yes& no& yes& $\geq$ 1 & polygon\\
         \hline
         Proposed method & \textbf{computed}& \textbf{yes}& \textbf{yes}& \textbf{yes}& \textbf{yes}& \textbf{yes} & \textbf{$\geq$ 1} & \textbf{polygon}\\
         \hline
    \end{tabular}}
     \smaller
    \caption{Related works that utilize similarity or adjacency to create treemaps. 
}
    \label{tab:table_hierachy_similarity}
\end{table*}

\section{Related Work}\label{sec:related}
For an overview of graph-based visualization methods for information visualization, please refer to~\cite{Gibson2012surveytwodimensional,Vehlow2016VisualizingGroupStructures}.
Here, we discuss various related treemap and treemap-like visualizations. 
Scheibel et al.~\cite{treemap_taxonomy} provide an extensive overview of treemap layout algorithms spanning from 1991 to 2019. 
The authors categorize them into different types of treemap structures. 
Following their nomenclature, we will mainly discuss compact space-filling treemaps.
Unlike node-link diagrams, treemaps visualize hierarchical structures by minimizing empty spaces, thus maximizing spatial efficiency. Treemaps additionally use the cell size to represent a data attribute. 

We will use the term treemap to describe a hierarchical diagram that exhibits the containment property, as defined in \cite{treemap_taxonomy}. In these diagrams, each child node is positioned and fully enclosed within the polygon of its parent node. In addition to the containment property, space-filling treemaps efficiently utilize space by recursively subdividing the bounding polygon to avoid any empty space. For rectangular treemaps, this bounding polygon is a rectangle that is further recursively subdivided into rectangles. 

Johnson and Shneiderman initially introduced rectangular treemaps in their works \cite{treemaps_orig, treemaps_orig2} to illustrate the hierarchical structure of file systems, with file size represented by the cell's area in the treemap. This use case remains prevalent, as seen in software like TreeSize \cite{TreeSize}. However, these early slice-and-dice techniques often yielded undesirable rectangle aspect ratios, leading to the development of various approaches \cite{Squarified_Treemaps, Ordered_treemap} to maintain aspect ratios closer to one.
In subsequent publications, a considerable design space of different treemaps and treemap-like visualizations was explored.  In \cite{Circular_Treemaps, ClockMap, bubble_treemap}, the rectangular structure was replaced by circular shapes, and \cite{origVoronoiTreemap, PowerHierarchy, FastVoronoiTreemap} use Voronoi diagrams. Space-filling curves, such as Hilbert, Gosper, and Moore curves, were used to place rectangles to preserve neighborhoods in dynamically updating treemaps \cite{takEnhancedSpatialStability2013, scheibelConstructingHierarchicalContinuity2023} and to generate regions \cite{Jigsaw, GosperMap}. 
In cartographic approaches, nodes may also utilize similarity or geographical features in the algorithm, and are visualized using an underlying graph~\cite{weiEfficientStableCircular2023} or linear programming layout algorithms\cite{nickelMulticriteriaOptimizationDynamic2022}.

Apart from merely altering the shapes of cells, treemap modifications have been proposed to smoothly show how hierarchies change over time, such as the evolution of codebases within large software packages. Therefore, different treemap layouts \cite{stableVoronoi, Stable_Treemaps_Moves, PowerHierarchy, evocells, sondagStableTreemapsLocal2018, faccinvernierStableGreedyInsertion2018, yamaguchiVisualizationDistributedProcesses2003} preserving the layout structure across multiple evolution steps were proposed. Vernier et al.~\cite{TimeTreemap} discuss several methods to generate rectangular treemaps for time-dependent data.
Treemaps were further extended to visualize additional information in hierarchical data. Görtler et al.~\cite{bubble_treemap} introduced a method to encode uncertainty values related to leaf nodes by modulating the edge of the circles representing the nodes using a sine wave. In \cite{ClockMap}, the circles representing cells of the diagram are used to represent individual clocks.

In the overview survey by Scheibel et al.~\cite{treemap_taxonomy}, only a few publications consider the similarity between nodes. In \cite{RecMap, IsoMatch, Grid_Layouts_Distance_Preservation}, the similarity among cells within a single-level hierarchy is expressed by positioning them in close proximity. Buchin et al.~\cite{similarity_treemaps} introduce a method to preserve neighborhoods within 2-level rectangular treemaps. Already by Wattenberg~\cite{TreeMapStockMarket}, a swapping strategy based on similarity is employed to express similarities in the treemap layout. Nmap \cite{Nmap} uses a slice and scale technique to subdivide a rectangle into smaller rectangles, utilizing similarity in the initialization step.

\subsection{Voronoi Diagrams}
Given a set of points $P= \{p_1, \dots, p_n\} \subset \mathbb{R}^2$, a Voronoi diagram partitions the plane into a set of cells $\mathcal{C}$. A cell $C(p_i)$ of point $p_i$ consists of all points in $\mathbb{R}^2$ for which there exists no closer other point $p_j \in P$.
Formally, a Voronoi diagram is defined as \cite{voronoi_book} :
$P= \{p_1, \dots, p_n\} \subset \mathbb{R}^2$, where $2 \leq n < \infty$ and $p_i \neq p_j$ for $i \neq j$. \\
We call the region given by 
\begin{equation}
C(p_i)=\{x \in \mathbb{R}^2 \mkern3mu | \mkern3mu  ||x-p_i|| \leq ||x-p_j|| \text{ for all } i \neq j\}
\end{equation}
the Voronoi polygon of $p_i$ and the set $\mathcal{C} = \{C(p_1), \dots, C(p_n)\}$ the Vornoi diagram of $P$.

To distribute the points more evenly, Lloyd's algorithm~\cite{Lloyd1982Leastsquaresquantization}, also called Voronoi relaxation, can be applied to generate an equidistant Voronoi diagram in which the points within $P$ are moved. This relaxation is calculated in the following way: For each iteration step, the centroid of each cell $C(p_i)$ is calculated, and $p_i$ is moved towards it, then the Voronoi diagram is recalculated. This procedure repeats until the points move less than a threshold value. A fully relaxed Voronoi diagram is also called a centroidal Voronoi tessellation (CVT).

To generate Voronoi diagrams with differently sized cells, a weight $w_i$ is assigned to each point $p_i$~\cite{origVoronoiTreemap}. The distance function is then adjusted to produce additively weighted power Voronoi tessellations characterized by polygon cells with straight edges.
\begin{equation}
C(p_i)=\{x \in \mathbb{R}^2 \mkern3mu | \mkern3mu  ||x-p_i||^2-w_i \leq ||x-p_j||^2-w_j \text{ for all }i \neq j\}
\end{equation}
Other weighting strategies exist. Multiplicatively weighted Voronoi tessellations generate hyperbolically curved edges.

\subsection{Voronoi Treemaps}
Voronoi treemaps, as described in \cite{origVoronoiTreemap}, are space-filling hierarchical Voronoi diagrams following the recursive subdivide structure of treemap generation techniques, as seen in~\cite{Squarified_Treemaps, treemaps_orig}. 
Voronoi treemaps exhibit a more organic appearance than their rectangular counterparts. Their convex-shaped bounding polygons make them less constrained than rectangular treemaps and applicable to a broad spectrum of domains. Unlike rectangular treemaps, where cells typically have at most four neighbors, they potentially adjoin more neighboring cells. This is advantageous 
when expressing node similarity via shared cell edges, as more node similarities can be preserved.

To construct a Voronoi treemap, a Voronoi diagram is generated for the first level of the tree hierarchy. 
The depth of the hierarchy defines the distance to the root node, while the height is the longest downward path to a leaf of the node, with leaf nodes having no child nodes.
Each cell in this diagram now serves as the bounding polygon for a new Voronoi diagram of its children. Thus, all diagrams on a level can be generated independently and in a top-down manner. 
A weighted Voronoi diagram is calculated and relaxed using an adaptation of Lloyd's algorithm~\cite{origVoronoiTreemap}. 
This optimization procedure adjusts the weights of the cells in each optimization step to reduce the error between a cell's actual and its desired area. Their computation was improved by~\cite{FastVoronoiTreemap}, and stable versions were proposed by~\cite{stableVoronoi, vanheesStablePredictableVoronoi2017, gotzDynamicVoronoiTreemaps2011}. Other optimization methods, such as L-BFGS for sites with Newton's method for weights, are also suitable~\cite{PowerHierarchy}.

\subsection{Similarity-preserving treemaps}

The survey by Scheibel et al.~\cite{treemap_taxonomy} provides a general distinction on whether treemap techniques use similarity or adjacency/neighborhood to create treemaps. However, a more fine-grained separation of these concepts is needed as the terms are ambiguous. In Section~\ref{sec:properties}, we, therefore, distinguish between both concepts and indicate when each is applied in the treemap visualization pipeline. 

In~\autoref{tab:table_hierachy_similarity}, we give an overview of previous work that incorporates similarity and adjacency for treemap visualizations and summarize other attributes of the approaches, such as how many levels of the hierarchy can be visualized and the shape of the treemap cells. 
Most approaches only utilize similarity to create and initialize the geometric cells or try to preserve already given layout neighborhoods without considering similarity.
During the subsequent Voronoi treemap optimization, the influence of the similarity-aware Voronoi cell initialization on the layout is not preserved, and cell neighborhoods may change.

From the publications surveyed by Scheibel et al.~\cite{treemap_taxonomy}, some mentioned techniques cannot be considered as multi-level treemaps, as they do not feature multiple hierarchy levels. Two of these techniques are RecMap~\cite{RecMap}, which uses non-overlapping rectangles arranged to abstract geographical entities and preserve their neighborhood relations, and IsoMatch~\cite{IsoMatch}, which arranges similar images in a grid or other shape and aims to preserve their pairwise similarities.
In their method, Buchin et al.~\cite{similarity_treemaps} build a rectangular 2-level treemap based on geographic positions. It aims to preserve the topology of geographic regions.
Ghoniem et al.~\cite{ghoniemWeightedMapsTreemap2015} propose Weighted Maps which also aim to preserve the relative position of geographic regions in rectangular treemaps with multi-level hierarchies.

Planar space-filling Gosper curves are used by Auber et al.~\cite{GosperMap} to place hexagons in the plane in the order of depth-first traversal of the hierarchy tree. Thus, the order of the nodes in the tree is preserved, and adjacent nodes in the hierarchy tree have adjacent cells in the GosperMap. However, only the adjacency of two nodes is considered due to the linear arrangement. Other similarity constraints, apart from the strict ordering of the nodes, cannot be preserved.

Nocaj and Brandes~\cite{Reference_Map} present, to our knowledge, the first Voronoi treemap incorporating similarities in the layout process. They use document similarities to create a 2D projection via MDS, which then initializes the layout of Voronoi diagrams to visualize queries in document collections. To fit the results in the respective Voronoi parent cell, they scale them. To calculate the similarities of documents, they use a vector-based TF-idf~\cite{Eck2009} approach with cosine distance. They note that determining the heuristic, dataset-specific threshold to create the similarity graph is tedious, as it has to be done manually.

Yao et al.~\cite{PowerHierarchy} propose a method to preserve the topological structure of a Voronoi treemap. It focuses on preserving the topology of dynamically changing Voronoi treemaps but does not use similarity to initialize the Voronoi cells. 
However, we argue that placing new nodes at the border of their new parent Voronoi cell or copying the positions of data nodes that change their parent in the hierarchy does not preserve the neighborhoods between cells.

Hawes et al. propose CodeSurveyor~\cite{hawesCodeSurveyorMappingLargescale2015} to visualize large hierarchical code bases using regions similar to islands partitioned into Voronoi cells. Using an energy minimization-based graph layout, similar nodes in the tree are placed in proximity and close to their parents.

Nmap~\cite{Nmap} uses a slice and scale technique to subdivide a rectangle into smaller rectangles while shifting their position to represent a weight property by the rectangles' area. The similarity is obtained in the initialization step by projecting the data points into a 2D bounding rectangle using MDS and only moving them by linear transformations, not affecting their similarity relationships inside a rectangle.

To summarize, although previous works utilize similarity in the initialization of treemaps or diagrams, no current method utilizes it during the final optimization step, where cells move.
We argue that changing only the nodes' initial positions is insufficient to preserve neighborhood relations in the final visualization. 
This can be observed in \autoref{tab:results}, where utilizing similarity only during initialization results in less preserved neighborhood relations.
Therefore, we additionally adapt the optimization step to influence the movement of cells to preserve the desired cell neighborhoods.

\section{Similarity \& Neighborhoods in Voronoi Treemaps}
\label{sec:properties}
Scheibel et al.~\cite{treemap_taxonomy} describe that treemap layout techniques may use attributes, such as weight, neighborhood, or similarity, during the layout process. However, in the above survey paper, no clear definition or distinction between similarity and neighborhood is given. 
Therefore, we clearly define the notions of
similarity and adjacency/neighborhood
in the context of treemaps. \\
In this paper, we use the term \textit{similarity} to describe features of the hierarchical input data. To describe geometric adjacency of neighboring cells in a treemap, we use the term \textit{neighborhood}. 
Preserving data similarity means that Voronoi cells are initialized and optimized to represent a node's similarity to other nodes.
\textit{Neighborhood-preserving} means that Voronoi cells representing similar nodes should have a shared neighborhood, i.e, share cell borders, and strive to retain this neighborhood during optimization. 
Furthermore, we contribute a more fine-grained analysis of previous work for space-filling treemaps and how and in which steps of the treemap visualization pipeline either notion is used (see Table~\ref{tab:table_hierachy_similarity}). 

\paragraph*{Similarity:}
We define a similarity feature $s$ as a $d$-dimensional normalized vector attribute $[0,1]^d$.
Words and their contextualized word embedding vectors, extracted from LLMs such as BERT~\cite{devlin-etal-2019-bert}, are an example of such a similarity feature.

Alternatively, the similarity attribute can be a co-occurrence vector of $d$ features encoded as a one-hot vector $\{0,1\}^d$ where 1 defines the occurrence of a feature. An example feature vector would be the genres of movies.
A particular case are data attributes with a topological meaning, e.g., geospatial neighborhoods like neighboring countries. Here, the similarity feature describes a geographic attribute of the data.

The similarity between nodes is the result of a similarity function applied to the similarity feature vectors.
\begin{equation}
f_{sim}: [0,1]^d \times [0,1]^d\rightarrow [0,1] \label{eq:sim_function}
\end{equation}
Different similarity functions might be appropriate depending on the kind of similarity feature that exists for the data.

\paragraph*{Geometric Adjacency/Neighborhood:}
In the literature, the term \enquote{neighborhood} is used in different contexts. It is used when referring to input data and its attributes, which describe their geographic arrangement, e.g., express that two countries share a border. 
However, it is also used when referring to the final geometric layout in a treemap, where adjacent cells share edges, meaning they have a shared neighborhood. 
Cells $c_1$ and $c_2$ are neighboring if $depth(c_1)==depth(c_2)$ and if any of the cell edges of $c_1$  and $c_2$ overlap.

\paragraph*{Initial Layout}
Many previous works do not aim to create an initial layout based on the data similarity but use an external layout as the basis of their method. The goal of these methods is to \textit{preserve already existing neighborhoods} in the initial, given layout. 
Instead, we propose a unified approach where data similarity is considered during all pipeline steps and neighborhoods are preserved during the Voronoi optimization without any external given initializations.
\begin{figure}[tb]
    \centering
    \vspace*{4.5mm}
    \includegraphics[width=\linewidth]{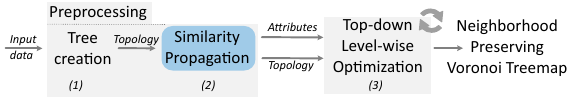}
    \caption[]{High-level overview of the pipeline to create neighborhood-preserving Voronoi treemaps, adapted from \cite{Scheibel2020TaxonomyTreemapVisualization}. Extensions are marked in blue. We (1) transform the hierarchical input data into a tree structure and introduce a (2) fully automatic similarity propagation step and (3) optimize the Voronoi diagrams on each level in a top-down manner.}\label{fig:exteded_treemap_pipeline}
    \vspace{-2mm}
\end{figure}

\section{Method}\label{sec:method}
In the following section, we give details of our proposed method.
We extend the existing treemap layout pipeline proposed by Scheibel et al.~\cite{Scheibel2020TaxonomyTreemapVisualization} at each step, utilizing similarity and cell neighborhood, as shown in Fig~\ref{fig:exteded_treemap_pipeline}.
We follow the example dataset shown in~\autoref{fig:teaser} and the respective optimization shown in~\autoref{fig:algorithm_example} concerning countries and their borders to explain the general procedure. 

\subsection{Overview}
Firstly, we provide a brief overview of the proposed system, as shown in ~\autoref{fig:exteded_treemap_pipeline}.
Given the input tree with a similarity feature vector on all leaf nodes, in a bottom-up preprocessing step, similarities are aggregated up the tree.
As in the worst case, all nodes are similar to each other, the complexity is reduced by binning and filtering the similarities of each node, extracting \textit{constraints}.
We utilize a breadth-first strategy to create Voronoi diagrams using Voronoi cells of the previous hierarchy level as borders. We start with the tree's \emph{root} as the topmost node, whose cell can have an arbitrary convex polygonal shape.

For every node, we create a new Voronoi diagram on the current level and initialize its geometry using the similarity constraints and a CVT matching step. 
After initialization, we swap the assignment of a node to a cell in the diagram to improve cell neighborhoods according to the constraints and optimize the Voronoi cells. 
All cells of all diagrams on a hierarchy level move to preserve their cell neighborhoods, get closer to their centroids, and grow the cells until their desired sizes are reached or the maximum iteration count is reached.
For the final treemap visualization, edges are visually altered to encode neighborhood relations using a symmetric \enquote{Interlocking} metaphor.

We implemented a preliminary version of our prototype\footnote{\url{https://graphics.uni-konstanz.de/voronoitreemap/}} in JavaScript and D3.js and provided the code\footnote{\url{https://github.com/cgmi/Neighborhood-Preserving-Voronoi-Treemaps}}
It shows projections of the tree's similarities and the optimization queue of the Voronoi treemap on each hierarchy level. 
The user can skip through the individual steps of the optimization to see how the cells move while preserving neighborhoods. 
We offer different interactions to investigate the cells' constraints.

\subsection{Preprocessing}\label{sec:preprocessing}
\begin{figure}[t]
    \centering
        \includegraphics[width=0.7\linewidth]{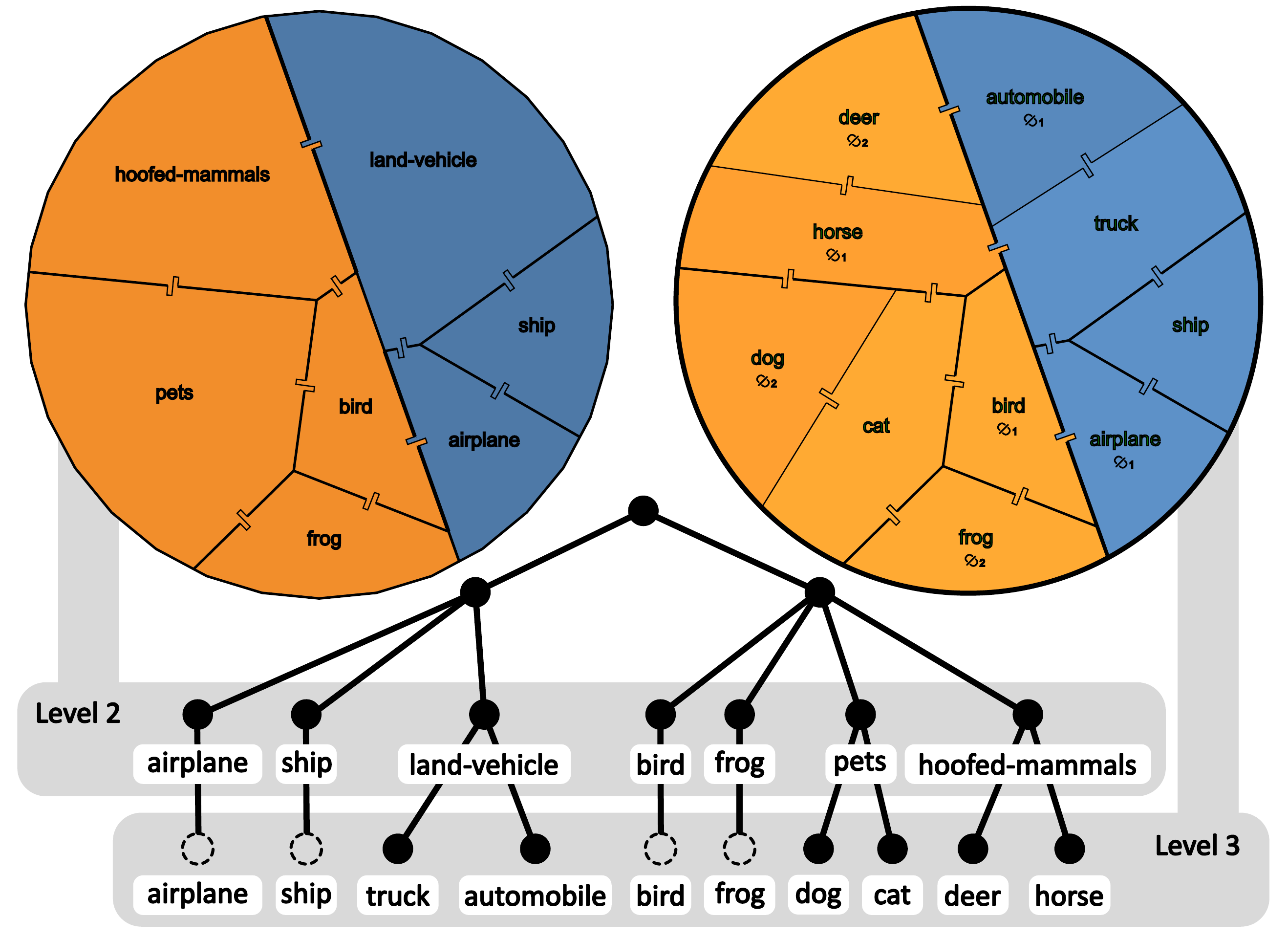}
            
         \caption{Cifar-10 dataset with virtual nodes: Top Left: Voronoi map of level 2, Top Right: Voronoi map of level 3. Virtual nodes (in grey) are needed for across-level relations, e.g., \emph{airplane} and \emph{automobile}. }
         \label{fig:overview_cifar}
\end{figure}

The input graph should be a tree, thus every node has exactly one parent, and a common root node exists.
To enable the calculation of similarity relations that cross hierarchy levels, we add artificial virtual nodes that propagate the node's data so that the depth is uniform. 
This can be observed in \autoref{fig:overview_cifar}, where nodes \textit{automobile} and \textit{airplane} are linked even though they are not on the same level in the hierarchy. \\
Our method takes as input a tree $T = (V,E)$, with nodes $V$ and edges $E$.
A weight $w$ is assigned to all leaf nodes. We use the notation $V^w$ to express the weight of node $V$.
Similarly, a $d$-dimensional similarity feature vector $s \in [0,1]^d$ is given for all leaf nodes of the tree. This similarity feature vector describes the nodes' properties, e.g., the neighboring countries in the \autoref{fig:teaser} or genres of a movie in \autoref{fig:imdb}. We use the notation $V^s$ to express the similarity feature vector $s$ of node $V$.

By propagating the similarities bottom up in the tree, we ensure that similarity is considered across all hierarchy levels.
The preprocessing and filtering step occurs directly on the given hierarchical data, usually in a tree structure.

\subsection{Similarity and Weight Propagation}
The treemap algorithm uses a top-down approach, so all nodes of the hierarchy from top to bottom must have a similarity and weight attribute assigned to them. 

\paragraph*{Leaf Node Aggregation}
Frequently, similarity information is only available on the leaf node level. In this case, we have to aggregate the values from the children to the parents. The similarity feature vector for node $V$ is defined as
\[
V^s =
\begin{cases}
V^s , & \text{if } V \text{ is a leaf}\\
\frac{1}{|children(v)|}\sum\limits_{V_{c} \in children(V) }V_c^s  , & \text{else}\\
\end{cases}
\]
In the \textit{Countries Dataset} (shown in \autoref{fig:teaser}), each leaf node corresponds to a distinct country whose similarity features describe its neighborhood status with other countries. To visualize this dataset, we must aggregate the similarity attribute up the hierarchical structure. \autoref{fig:teaser} shows \textit{China} and \textit{Russia}, which are on different continents but share a border. Therefore, to ensure the countries can have neighboring cells, their two parent nodes, i.e., \textit{Europe} and \textit{Asia}, must also share a Voronoi edge in the hierarchy level above.
This concept aligns with the principle of upward uncertainty aggregation described in \cite{bubble_treemap}.

\paragraph*{Weight Aggregation}
Weights are, like the similarity feature vectors, also aggregated in a bottom-up manner. But instead of averaging the embedding vectors, we add up the weight of the child nodes to determine a parent's weight value:

\[
V^w = 
\begin{cases}
V^w , & \text{if } V \text{ is a leaf}\\
\sum\limits_{V_{c} \in children(V) }V_c^w  , & \text{else}\\
\end{cases}
\]
The output of the preprocessing step is the expanded tree topology, with the similarity attribute $V^s$ and the weight attribute $V^w$ on each node $V$.
\subsection{Similarity Binning \& Filtering}

Starting from the leaf nodes, we calculate the similarities between all nodes on a level. $f_{sim}$ is a dataset-specific similarity function, such as cosine similarity or Jaccard coefficient~\cite{Eck2009}. 
The user can set it to match the dataset.

Ideally, we want to keep all similarities between all nodes; however, for most real-life datasets, this would result in a fully-connected graph, if we define similarities as edges and nodes as vertices.
Creating a neighborhood-preserving Voronoi treemap is related to finding a planar embedding of this graph. A subset of edges has to be selected that maximizes the similarity retained, which is NP-hard~\cite{michael1979garey}. 
If all nodes are highly connected to one another, it is generally impossible to realize all similarities as cell neighborhoods in the Voronoi treemap. Due to its hierarchical nature, our case is even more difficult, as each cell has to be positioned within its parent cell, and each cell has to keep a specific distance according to its weight attribute. Additionally, our neighborhood-preserving layout has to match the subset of edges to actual Voronoi cell neighborhoods on the geometric level.
We further discuss the complexity of layouting this graph in \autoref{sec:discussion}.

Therefore, we propose reducing the complexity initially, retaining only the strongest similarities between nodes as \textit{constraints} using a binning function. The details for the binning can be found in the supplementary material.
We use five bins distributed from 1 to 0, and assign the similarities of each node to all other nodes accordingly.
We iterate through all bins and stop when we find the first empty one. 
The nodes in the bins up to that point form our \textit{constraints}.
Constraints between nodes are undirected.

\begin{figure}[tb]
    \centering
    \includegraphics[width=\linewidth]{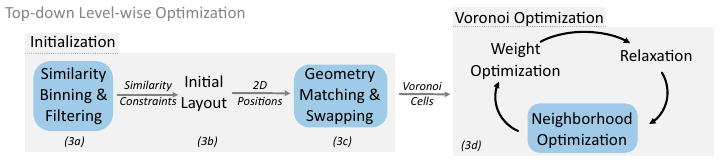}
    \caption[]{Level-wise optimization: For each Voronoi diagram on the same level, with our proposed extensions marked in blue: (3a) Bin and filter node similarities, (3b) calculate initial layout, (3c) match each node to a Voronoi cell. (3d) The cells of each Voronoi diagram are optimized to move towards their centroid while optimizing cell neighborhoods according to the similarity, and grow cells to match the node's weight.}
    \label{fig:treemap_pipeline_level}
\end{figure}

\subsection{Top-Down Treemap Queue}

\begin{figure*}[t]
\vspace{-1cm}
    \centering
    \begin{subfigure}[b]{0.13\textwidth}
        \includegraphics[width=\linewidth]{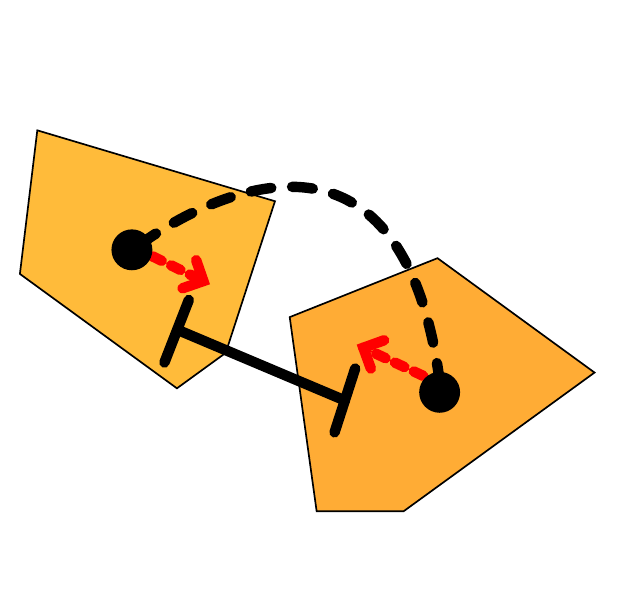}
        \caption{}\label{fig:attraction_a}
    \end{subfigure}
    \begin{subfigure}[b]{0.13\textwidth}
        \includegraphics[width=\linewidth]{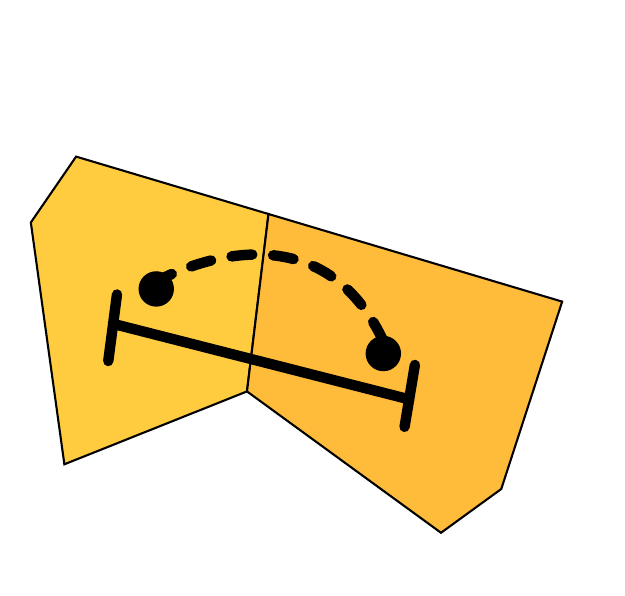}
        \caption{}\label{fig:attraction_b}
    \end{subfigure}
    \begin{subfigure}[b]{0.13\textwidth}
        \includegraphics[width=\linewidth]{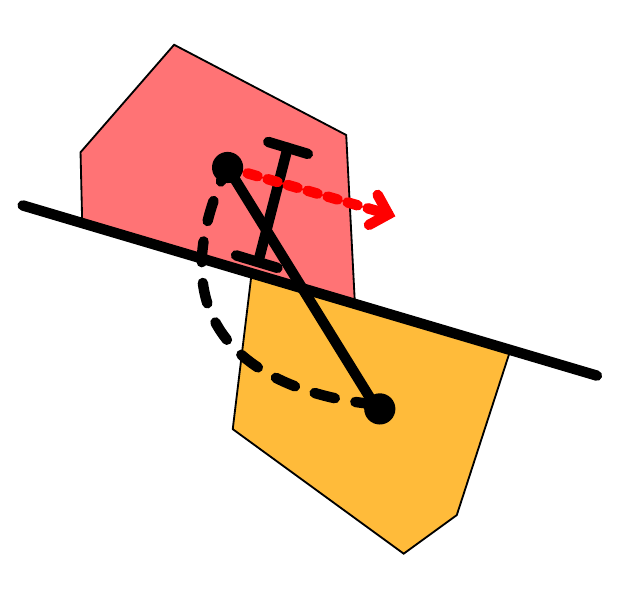}
        \caption{}\label{fig:attraction_c}
    \end{subfigure}
        \begin{subfigure}[b]{0.13\textwidth}
        \includegraphics[width=\linewidth]{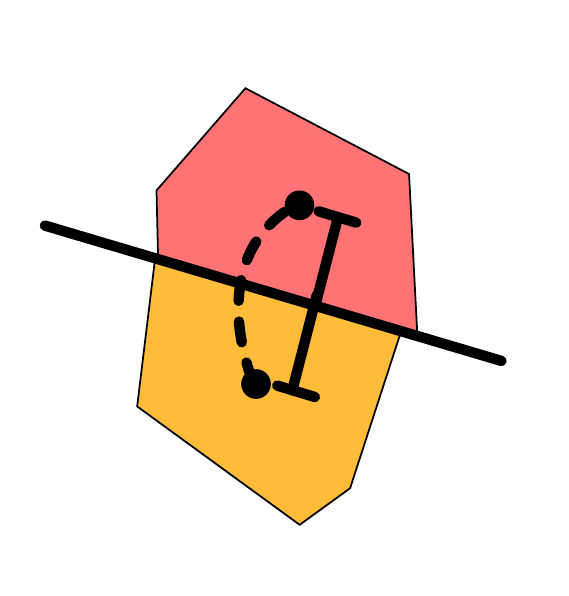}
       \caption{}\label{fig:attraction_d}
    \end{subfigure}
            \begin{subfigure}[b]{0.17\textwidth}
        \includegraphics[width=\linewidth]{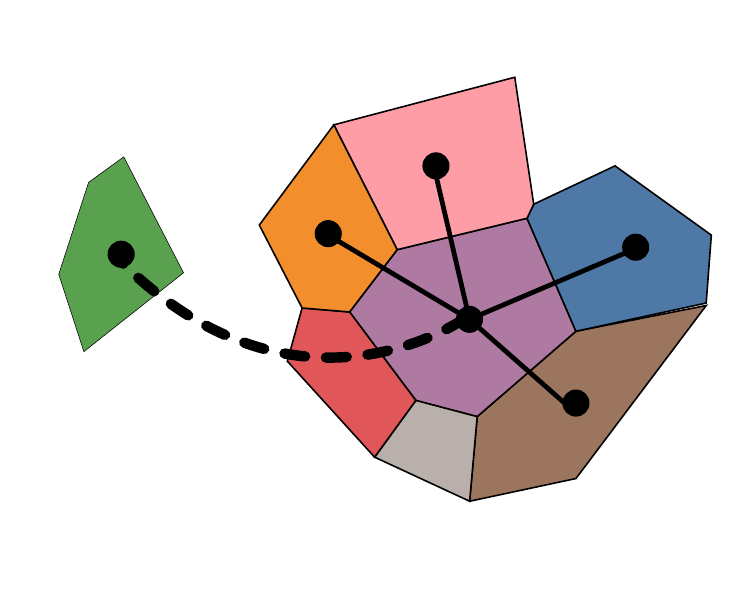}
        \caption{}\label{fig:attraction_e}
    \end{subfigure}
     \begin{subfigure}[b]{0.13\textwidth}
        \includegraphics[width=\linewidth]{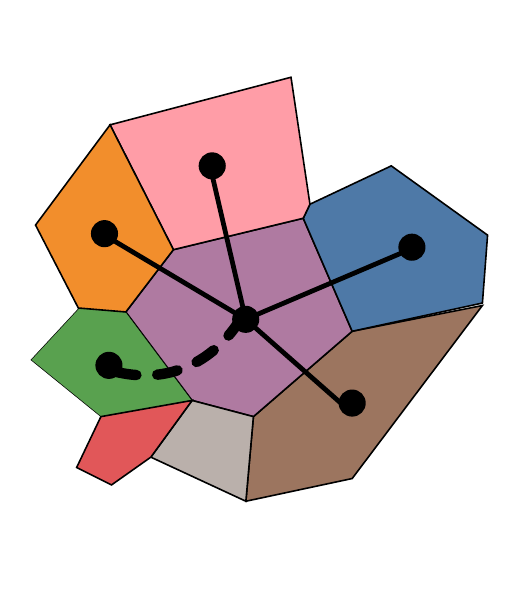}
        \caption{}\label{fig:attraction_f}
    \end{subfigure}
    \vspace{-2mm}
    \caption[]{Overview of different neighborhood scenarios: Two cells connected via similarity constraints (represented via the dashed line) move towards each if they are not neighbors and the distance is larger than $distance_{max}$~(a). This continues until the cells reach a $distance_{min}$ (b). If the link crosses parent cell edges and the distance is between $distance_{min}$ and $distance_{max}$, we move cells orthogonal to the parent edge (c) to optimize positioning to align both cells (d). If cells are not neighbors, but share a constraint (other constraints as black lines), we only move towards it if some neighbors do not have constraints with the target cell (e) to create the neighborhood~(f).}\label{fig:attractionExamples}
\end{figure*}
After preprocessing and extracting the constraints, we create the neighborhood-preserving Voronoi treemap.
Each level of the tree is realized as one or multiple Voronoi diagrams in the Voronoi treemap.
As seen in \autoref{fig:treemap_pipeline_level}, we first initialize the diagrams on each hierarchy level and then subsequently optimize them.
A top-down, breadth-first strategy is used to iterate the tree and populate a queue with the respective Voronoi diagram initializations.

We process the queue level-wise and extract all nodes belonging to the current hierarchy level.
In the countries dataset (shown in \autoref{fig:teaser}), for example, we extract the first group of nodes from the queue, which contains only the $\{root\}$ node.
We obtain all the child nodes $\{V_{1}, \dots V_{n}\}$ of the root node; these would be the \textit{continents}.
On the next level, we extract for each $\{V_{1}, \dots V_{n}\}$ its children.  We have, therefore, groups of children \{$\{V_{1_1},  V_{1_2}, \dots\}, \{V_{2_1}, V_{2_2}, \dots\}, \dots, \{V_{n_1}, \dots, V_{n_m}\}\}$ as a result.
In our example, these would be the \textit{continents} and their respective \textit{countries}.
Each of these nodes will create a Voronoi diagram.
The outer shape of the Voronoi diagram is given by the user for the root node, or it is the cell shape of the parent Voronoi cell for all other nodes.
After all the diagrams on the current level have been successfully initialized and the Voronoi geometry has been created, we optimize them. As we aim to preserve the cell neighborhoods, all Voronoi diagrams on a level have to be optimized jointly.
It is equally important that every map performs an optimization step in a round-robin fashion. 
This way, the updated position can influence the iterative optimizations of the other Voronoi diagrams on the same level.
After traversing all the levels above the leaf nodes, the queue is empty, and we render the final neighborhood-preserving Voronoi treemap.

\begin{algorithm}[t]
    \DontPrintSemicolon{}%
    \SetKwFunction{Optimization}{Optimization}%
    \SetKwProg{fn}{function}{}{}%
    \fn{\Optimization{$ cells $}}{
    $iterCount$ $\gets$\ 0\;
      \While{iterCount < maxIterCount}{  \label{algo:outer:maxIteration}
          $updateNeighborhood(cells)$\;\label{algo:outer:updateneighbor}
            \ForAll{$c$ of $cells$}{
                 NeighborhoodOptimizationStep($c$)\label{algo:outer:neighborstep}
             }
            
             \If{iterCount $\geq$ $0.8$ maxIterCount}{\label{algo:outer:weight}
               WeightOptimization($cells$)\label{algo:outer:weightstep}
                }
             iterCount++\;
       }
     }
    \caption{Neighborhood-preserving Voronoi Treemap}
  \label{alg:VoronoiTreemapAlgorithm}
\end{algorithm}

\begin{algorithm}[t]
    \DontPrintSemicolon{}%
    \SetKwFunction{NeighborhoodOptimizationStep}{NeighborhoodOptimizationStep}%
    \SetKwProg{fn}{function}{}{}%
    \fn{\NeighborhoodOptimizationStep{$ c $}}{
                $constraints \gets$\ Get constraints of $c$\;\label{algo:inner:getEdges}
                $sort(constraints)$\;\label{algo:inner:sortEdges}
                didMove $\gets$\ $false$\;

                    \ForAll{$constraint$ of $constraints$}{
                        c2 $\gets $Get other cell of $constraint$ \; 
                        \If{c2.simNeighbors<maxNeighborCount}{\label{algo:inner:maxConstraints}

                            \If{!neighbors(c, c2)}{\label{algo:inner:neighbors}
                                c.moveToward(c2)\;
                                didMove $\gets$\ $true$\;
    
                            }
                            \ElseIf{!aligned(c, c2) and (c.parent != c2.parent)}{\label{algo:inner:aligned}

                                c.MoveOrthogonalTo(c2) \;
                                didMove $\gets$\ $true$\;
                            }
                        }
                    }
                    \If{!didMove}{
                        c.moveTowardCentroid()\;\label{algo:inner:centroid}
                    }
    }
    \caption{Neighbor Optimization Step}
    \label{alg:VoronoiTreemapAlgorithm2}
\end{algorithm}

\subsubsection*{Geometric Initialization}

Cells can't be positioned arbitrarily, but must lie within their parent Voronoi cell.
This layout step is performed on each hierarchy level before creating the Voronoi diagrams. 
The computed 2D positions of the nodes are then used as initial positions for the Voronoi cells.

Many previous works~\cite{Nmap, IsoMatch, Reference_Map} use a \textit{dimensionality reduction} technique such as MDS, PCA, UMAP~\cite{McInnes2018}, t-SNE~\cite{vandermaaten08a}, or a force-directed approach~\cite{abuthawabeh2020force} to create 2D positions for all nodes on the leaf level. 
These methods, however, typically lack the ability to directly confine the placement of nodes within a predefined boundary shape. This limitation may not pose a significant problem when dealing with a single hierarchy level. In scenarios where multiple diagrams coexist on the same level, the disparity between the shape of the dimensionality-reduced point cloud and that of the boundary polygon might diverge significantly. Addressing this challenge is crucial, necessitating strategies to restrict node positions post-dimensionality reduction.

One approach to mitigate this issue involves scaling all nodes grouped by the hierarchy (sibling points) to ensure they fit within their parent boundary shape~\cite{Reference_Map}. Nevertheless, this solution may not be effective in the presence of outliers, as the scaling could result in points being positioned very close to one another. An alternative is to place cells that lie outside their parent in the nearest position along a ray cast from the cell center to the position~\cite{PowerHierarchy}, or utilize Wachpress coordinates~\cite{stableVoronoi}.\\

\subsubsection*{Matching }

In contrast to previous methods, we use a similarity-based matching utilizing the Kuhn–Munkres algorithm~\cite{Munkres1957AlgorithmsAssignmentTransportation} that takes the Euclidean distance between cell centroids into account. All nodes are matched to cells of a fully optimized CVT, as this improves the optimization process. Similar to Yao et al.~\cite{PowerHierarchy}, we use a random initialization of the CVT.
Then, we match cells and nodes 
in a way that most effectively preserves the distances observed in the high-dimensional space.
As we can see in \autoref{fig:algorithm_example}, the initialization step already preserves the neighborhood between the continents of \textit{Asia} and \textit{Europe}.
To improve the neighborhoods further, we \textit{swap} the assignment of nodes to Voronoi cells if this realizes more constraints as cell neighbors.
Thus, by swapping, in a well-behaved dataset such as the teaser~(\autoref{fig:teaser}), we can increase the neighborhood preservation to 83,33\% of the similarity constraints (\autoref{tab:results}). Across all datasets, the avg. preserved constraints after are 62,45\%, compared to random CVT(avg. 34,63\%) and proj.+scale(avg. 39,3\%).
\subsubsection*{Neighborhood-preserving Voronoi Optimization}

\begin{figure}
    \centering
\includegraphics[width=\linewidth]{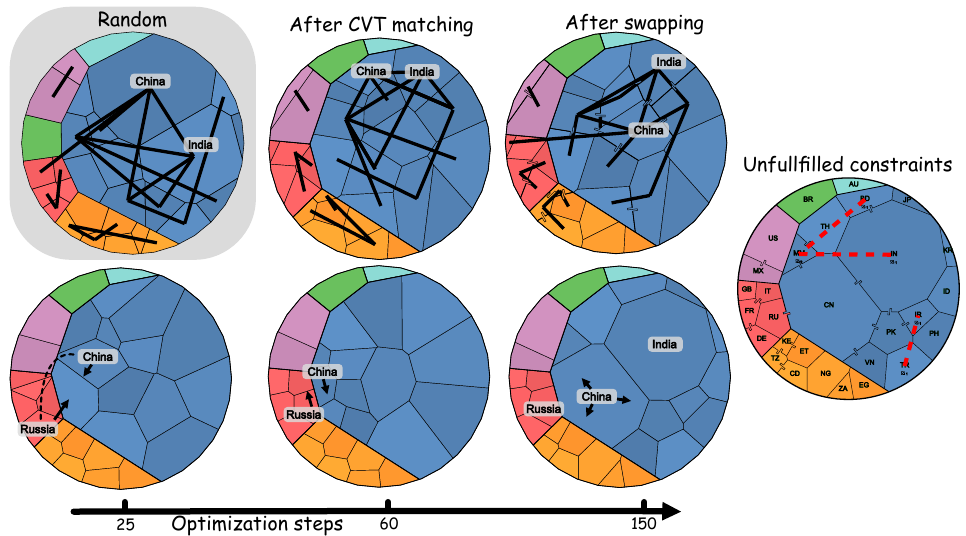}
     \caption{Left: Example optimization run: Nodes are matched to the cells of a CVT based on similarity, then we swap cells to improve neighborhoods. Black lines show the similarity constraints. Then, the diagrams on each level are optimized to preserve neighborhoods. In the last 20\% of the iterations, cells grow to match their weight attribute.
     Right: Unrealized Constraints are indicated by dashed red links.}
         \label{fig:algorithm_example}
\end{figure}

During the optimization step, where cells move towards their centroid and grow to match their weight attribute, a crucial component is maintaining the previously constructed Voronoi cell neighborhoods.
Therefore, we propose an additional similarity force during the optimization step that attracts cells that are connected via a \textit{constraint}.
We start processing the optimization queue (\autoref{alg:VoronoiTreemapAlgorithm}), where, in a round-robin fashion, the diagrams on each level perform an optimization step. 
This consists of (1) calculating the cell neighborhoods~(\autoref{algo:outer:updateneighbor}), (2) moving the cells~(\autoref{algo:outer:neighborstep}), and (3) growing the cells~(\autoref{algo:outer:weightstep}). 
Because the movement of cells of one diagram influences the cell neighbors of all other diagrams on the same level, cell neighbors must be recalculated in each step.
In the traditional Voronoi diagram relaxation, cells iteratively move toward their centroid to generate visually pleasing cells. 
However, this movement is independent of the similarity between nodes.
Therefore, every movement can potentially decrease the preserved neighborhood relations of the treemap.
We thus directly influence the movement of Voronoi cells during optimization so that cell neighborhoods are preserved.

In a given optimization step of a Voronoi diagram, for all nodes $V$ and their respective cells $c$, we update the cell neighbors(~\autoref{algo:outer:updateneighbor}), as a previous optimization step might have changed them. 
Iterating through all edges of a cell $c$, we assign each edge of $c$ a neighboring cell $c_i$.
Important to note here is that these neighborhoods may cross treemap hierarchy borders.
For each cell $c$, we then perform a neighborhood optimization step~(\autoref{algo:outer:neighborstep}).

In \autoref{alg:VoronoiTreemapAlgorithm2} we first extract all constraints of the current cell $c$~(\autoref{algo:inner:getEdges}).
We sort the similarity constraints $[s_{(v_1,v_2)},...,s_{(v_n,v_m)}]$ of the current cell
by (1) their sim. function, defined in~\autoref{eq:sim_function}~(\autoref{algo:inner:sortEdges}), and further by their euclidean distance.
The constraint whose cells are furthest away from each other in the diagram is processed first. 
This enables the corresponding cells to move towards each other even if they may have weaker similarity constraints with their currently surrounding cells.
For each constraint $s$, we obtain the connecting target cell $c2$. 
We determine the neighborhood status between $c$ and $c2$~(\autoref{algo:inner:neighbors}).
If $c$ and $c2$ do have a constraint, and are not neighbors, we move $c$ straight towards $c2$, until they either become neighbors or the cell edge in the direction to $c2$ overlaps with its parent Voronoi outer cell edge. 
The distance between cell nodes sharing a similarity constraint should be in the range of $distance_{min}$ and $distance_{max}$.
Once $c$ and $c2$ have moved to be within that range, and have the same parent, the similarity constraint is considered fulfilled. If they are cousin nodes, and their parents differ, another movement is needed. If cousin nodes share a common collinear cell edge but are not aligned along this edge~(\autoref{algo:inner:aligned}), we move $c$ orthogonal along the collinear cell edge towards $c2$.
However, if a cell $c$ has many constraints, oftentimes, not all of them can be realized as cell neighborhoods, as this can negatively impact the shape of the cells and the optimization convergence.
If the number of cell neighbors of node $c2$ is higher than the \textit{maximum \# neighbors}, we check if all the neighbors of $c2$ have constraints to $c2$.~(see \autoref{fig:attraction_e})
Only if some of the cell neighbors do not share similarity constraints, we move $c$ towards $c2$~(see \autoref{fig:attraction_f}).
The \textit{maximum \# neighbors} is a heuristic value, which was chosen so that area proportions of the cells do not degrade too much. 
If, due to the neighborhood criteria, we cannot move the cell $c$ to improve its neighborhood, we move it towards the polygon centroid~(\autoref{algo:inner:centroid}). We chose the Lloyd relaxation~\cite{Lloyd1982Leastsquaresquantization} for the Voronoi optimization due to its simplicity and readily available implementations. Other relaxation techniques such as~\cite{PowerHierarchy} could also be utilized.

An example of several subsequent optimization steps can be observed in ~\autoref{fig:algorithm_example}.
Nodes \textit{China} and \textit{Russia} are not neighbors, while they are connected by the strongest constraint.
We can see that \textit{China}'s cell slowly moves toward \textit{Russia} throughout multiple optimization steps. 
Russia's cell, however, cannot move further towards China, as it is already at the edge of its parent Voronoi cell.

As shown in~\autoref{fig:algorithm_example}, cells do not grow during the initial optimization steps. 
This is because an uneven distribution of the weights disrupts the free movement of cells inside their parents' borders. 
Therefore, we only grow cells during the last 20\% of the optimization steps~(\autoref{algo:outer:weight}).
After \textit{China} and \textit{Russia} moved closer, they started sharing an edge in iteration 60.
As can be observed, after 80 \% of the iterations, cells grow proportionate to their weights. 
Subsequently, we iterate through all the cells of a diagram and then move on to the following diagram until every diagram has performed an optimization step.
This continues until all diagrams have reached the maximum iteration count~(\autoref{algo:outer:maxIteration}).
Once all the diagrams of a level have finished, we continue to the next level until we reach the second-to-last level.

After this level is fully optimized, the final Voronoi treemap can be visualized.
Constraints that could not be realized, as seen in~\autoref{fig:algorithm_example}, may be highlighted using dashed lines.

\subsection{Voronoi Treemap Visualization}
After completion, when the Voronoi treemap queue is empty, we create the final visualization 

as shown in~\autoref{fig:teaser}.
The colors of the Voronoi cells are defined as:
\[
color(V) =
\begin{cases}
color(V), & \text{if } color(V)\\
tableau20(V), & \text{else if } depth(V)==1\\
color(V_{parent})+offset  , & \text{else}\\
\end{cases}
\]
If a node $V$ has a color attribute in the dataset, we utilize it. Otherwise, if $depth(V)==1$, the node is assigned one color from the Tableau10 or Tableau20 color schema, depending on the number of nodes with depth 1. Every node $V$ with $depth > 1$ uses a variation of their parents' color, with a small random offset in brightness and value.
The depth of the Voronoi diagrams in the hierarchy is represented by adjusting the stroke width of the cell outline, emphasizing the hierarchical structure.

Methods like \cite{tennekesTreeColorsColor2014, mertzQualityApproachHierarchical2024} could also be employed to additionally encode and highlight the depth using color.

 \subsection{Visual Encoding of Neighborhood Relations}
After the optimization, if we look at the neighborhood-preserving Voronoi treemap, we cannot tell whether or not Voronoi cells are similar to each other, because the similarity is not visually encoded in the diagram yet. 
We want to visually communicate to the reader whether an edge represents a similarity constraint between cells or if it is an edge that does not indicate a neighborhood relationship. 
As we already utilize the stroke width to encode hierarchy depth, we cannot simply adjust the stroke width to encode such information.
Instead, we encode the similarity between neighboring cells using an \textit{Interlocking} metaphor.
If two cells in a Voronoi diagram are neighbors due to a constraint, their path is modulated to include an interlocking symbol. 
We propose this symmetric interlocking metaphor as it conveys the notion of things belonging together. We use three different sizes for the interlocking symbols to express the similarity between the neighboring cells. In \autoref{fig:flare}, the cells in the enlarged cut-out region have different similarities, represented by the interlocking symbol length.
This allows the user to immediately see which cells are connected and which not. 
In the teaser, the relationship between China and Russia is especially visible, as their cells have different colors, 
and the path's intrusion into the other cells pops up.
We did not consider alternative ways to directly encode the similarity on the respective edges of the Voronoi cells, e.g., by varying the stroke width, as stroke width is already utilized to represent the hierarchical structure. Further, only adapting some cell edges of a cell breaks the uniformity of the cell outline.

\subsection{Interactions}
More information is available using the interactive interface:
To demonstrate the algorithm's iterative procedure, we implemented an interactive prototype with several common interactions. After selecting a dataset, we show a baseline classic Voronoi treemap and our neighborhood-preserving Voronoi treemap. Below, each step of the optimization loop of our proposed method is shown in a slider, allowing for detailed investigation. 
Once the user hovers over a cell, its similarity constraints are highlighted and the current Voronoi cell neighbors are displayed. To distinguish between neighboring cells linked by constraints and those not linked, we gray out all other cells that are not neighbors and outline cells connected via constraints with a stronger stroke-width.
We also include the option to visually show that some constraints were not realized, as their nodes are not neighbors. Cell neighborhoods sometimes cannot be preserved, for example, if their parents' nodes are too far away or other sibling nodes have a significant weight imbalance. In this case, the cell can include the \includegraphics[scale=0.6]{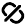} icon, to highlight the disconnectivity as shown in \autoref{fig:algorithm_example}. However, as this may be distracting, we only show this icon via a checkbox.

\section{Evaluation}
\label{sec:evaluation}
We evaluate the neighborhood-preserving Voronoi treemaps using four quantitative measures. 
For the two related works that are most relevant to our approach \cite{Reference_Map, PowerHierarchy}, we requested code from the authors but got either no response or the code could not be shared. Therefore, we carefully analyzed the paper to determine where similarity or neighborhood-preservation is used in the approach. Especially for~\cite{PowerHierarchy}, the topology preservation aspect is not described in detail in the paper, which makes a fair comparison difficult. We therefore only reimplemented the parts of these methods that concern similarity or neighborhood-preservation. To limit the scope of this paper, we do not compare the classic Voronoi optimization to other optimization methods.

For Nocaj and Brandes's paper, the authors precompute a so-called Reference Map, which is a Voronoi treemap. As they use a tf-idf approach to create a similarity vector for the documents, their data is similar to our CIFAR10 dataset. They also create a similarity graph based on a heuristic, dataset-specific similarity cut-off. During the layout step, they use MDS on the vectors and scale the 2D positions so that all nodes of a parent fit within the parent's Voronoi cell. However, we cannot precisely replicate their result, as we don't know which MDS implementation and what scaling factor were used. 
In their next step, dots, representing queries, are overlaid on top of the Reference treemap according to the similarity between the queries. These similarities can exist between nodes of different parents. 
Although the dots are placed according to the similarity of the queries, the positions are not used to create Voronoi treemap cells or another Voronoi treemap level. Therefore as they are not bound by the same restriction as Voronoi diagrams and can only be used for leaf nodes, it is not directly comparable to our method.

For Yao et al.~\cite{PowerHierarchy}, we observed that similarity is not considered during the layout step. The method solely aims to preserve topology during the geometry initialization if nodes change parents or new nodes are added in the next time step. This preservation is only shown for an example where nodes are attached to a parent node without children by scaling them to fit inside their new parent Voronoi cell. Although the authors propose a new method of relaxing and growing the cells of a Voronoi diagram, directly comparing this part of their method to ours is not useful, as they do not consider similarity or neighborhood preservation. Our relaxation could theoretically be switched out for theirs; we propose an additional step that can be used in conjunction with the relaxation and cell-growing step.\\
In summary, we reimplemented the  MDS and scaling of child nodes into their parent Voronoi cells, as proposed by Nocaj and Brandes~\cite{Reference_Map}. Then, a traditional Voronoi Cell initialization and optimization using Lloyd is performed. 
To compare ourselves to Yao et al.~\cite{PowerHierarchy}, we reimplemented the geometric initialization on a CVT with random assignment of the data points.
For the next time step they use the previous positions with new weights. As we do not use dynamic data, this is not applicable to our approach. 
\\
For quantitative evaluation, we calculate standard quantitative treemap measures, such as convergence error and aspect ratio of the bounding rectangles of the leaf nodes.
As proposed by previous work~\cite{Reference_Map}, we count the number of direct neighbors with constraints, as these should share cell edges. 
However, we argue that the direct neighborhood measurement alone is not expressive enough to convey how well neighborhood relations have been preserved.
Therefore, we propose two additional quantitative quality measures, the \textit{median shortest path distance}, and the \textit{maximum path distance} of all constraints between leaf nodes.
We calculate the shortest path for each constraint using the A* algorithm~\cite{Hart1968FormalBasisHeuristic} on the neighborhood/Delauney graph of the leaf nodes, and calculate the median path distance, the edge value is one.
This way, we can measure how far nodes that should be neighbors but share no edge have drifted, and median/max distance of these nodes. 
In cases where the weight attribute is highly unbalanced or multiple nodes are highly connected to each other, it is impossible to realize all their constraints as cell edges, because the Voronoi treemap would degrade too much. 
However, we can still move them close together and minimize their distance.

\begin{table*}[tb]
    %\tiny
    \centering
    \smaller
    \caption{Overview of the used datasets. We show different data characteristics, as well as neighborhood preservation metrics. The maximum iteration count of the Voronoi diagram optimization was 150. For each dataset, the result with the most realized similarity constraints as cell neighbors per dataset is in \textbf{bold}.}
    \resizebox{\textwidth}{!}{
    \begin{tabular}{|c|c|c|c|c|c|c|c|c|c|c|c|}
        \hline
        \multirow{2}{*}{Dataset} &
        \multirow{2}{*}{\thead{\# Total \\ Levels}} &
        \multirow{2}{*}{\thead{\# Total \\ Nodes}} &
        \multirow{2}{*}{\thead{\# Leaf \\Nodes}} &
        \multirow{2}{*}{\thead{\# Constraints\\ (leaves)}} &
        \multirow{2}{*}{\thead{Avg.\\Area \\Error}} &
        \multirow{2}{*}{\thead{Max \\Graph \\Dist.}} &
        \multirow{2}{*}{\thead{Avg.\\ Aspect \\Ratio}} &
        \multicolumn{4}{c|}{\# Constraints preserved (lowest level)} \\
        & & & & & & & %\multicolumn{5}{|c|}{}  
        & \thead{Random CVT \\ init \cite{PowerHierarchy}} 
        & \thead{Proj. + scale \\ init \cite{Reference_Map}} 
        & \thead{Match + swap \\ (our   method)} 
        & \thead{Neighbor optim. \\ (our  method)}
        \\
        % & \thead{max graph distance \\between constraints}\\
        \hline
        Country Pop.  & 3 & 36 & 29& 18& 0.04& 2& 1.16& 5(27.77\%) &11 (61.11 \%) & \textbf{15(83.33\%)} &\textbf{ 15(83.33\%)}\\
        \hline
        Cifar-10            & 4 & 20 & 10& 19 & 0.00 & 2 & 1.11& 9  (47.36\%) & 10 (52.63\%)& \textbf{14 (73.68\%)} & \textbf{14  (73.68\%)}\\
         \hline
         Flare  &  5  & 524  & 220 & 547 & 0.00& 14&1.03& 147 (26.34\%)& 151 (27.06\%)& \textbf{223 (40,76\%)}&  220 (40.21\%)\\
         \hline\
          SP500 Variance  &  6  & 655  & 429 & 1798 &0.02 & 21& 1.03&  103 (5.72\%)& 112 (6.22\%)& 178 (9.89\%)&  \textbf{179 (9.95\%)}\\
         \hline\
          IMDB Genres  &  3  & 80  & 59 & 42 &0.00 & 8& 1.02& 16 (23.88\%)& 14 (20.89\%)& \textbf{30 (71,42\%)}&  29 (69.04\%)\\
         \hline
        Synthetic m-n       & 2 & 11 & 10 & 11 & 0.00& 2 &1.10&  4 (36.36\%) & 4 (36.36\%)&\textbf{10 (90,9\%)} & 9 (81.8\%)\\
         \hline
        Synthetic 2 lvl  & 3 & 23 & 17 & 24 &0.00& 3& 0.96&  18 (75\%) & 17 (70.83\%) & \textbf{19 (79.16\%)}& \textbf{19 (79.16\%)}\\
         \hline
    \end{tabular}}
    \label{tab:results}
\end{table*}
\section{Results}\label{sec:results}
To showcase our method's usability, we created neighborhood-preserving Voronoi Treemaps of various datasets. 
We created a prototype of our proposed method in JavaScript. We use the D3 Voronoi Map~\cite{d3_voronoi_map} as a basis but extend it with our implementation. 
As can be seen, our method can create diagrams with a large number of nodes (>500), leaf nodes (>400), constraints (>1500), and hierarchy levels (>5).
However, with increasing hierarchy levels and constraints, the number of preserved neighborhoods decreases significantly, as movement on the lower level is more restricted. Our method, therefore, works best for hierarchies of height 2 and 3.
The complexity of our method is mainly limited by the height of the hierarchy, as well as the number of leaf nodes and the cost of the matching step utilizing the Munkres algorithm.
The calculation of the neighborhood relations is also costly, as all edges of the polygons within a level must be compared.
We estimate the overall complexity to be $d * n^3$, with $d$ being the depth of the tree and $n$ being the number of leaf nodes.
The Voronoi treemap can be computed within a few seconds per hierarchy level. 
In the prototype, all the optimization steps can be explored.

\subsection{Synthetic Datasets}
To investigate how our method handles edge cases, we created several synthetic datasets that cover a variety of different connectivity levels of the similarity graph, from very loosely connected to fully connected.
In the Appendix, we show how the optimization handles different edge cases.
The results can be observed in \autoref{tab:results}.
%APENDIX
\captionsetup[subfigure]{font=small}

\subsection{Real-World Datasets}
In~\autoref{tab:results}, we give an overview of properties and attributes of the datasets used in this paper. 
We selected a mixture of established and novel hierarchical datasets with meaningful semantic information available. The datasets are from different domains, covering linguistics, infographics, and economic data.
We chose the respective datasets because they contain inherently interesting, similar information, such as the semantic similarity of languages, whether or not countries share a border, or the similarity of stock movement across time.
\begin{figure*}[t]
    \centering
    \begin{subfigure}[b]{0.29\linewidth}
    \includegraphics[width=\linewidth]{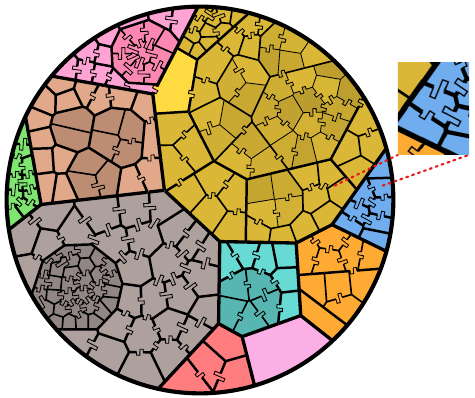}
     \caption{}
         \label{fig:flare}
\end{subfigure}
\begin{subfigure}[b]{0.25\linewidth}
    \includegraphics[width=\linewidth]{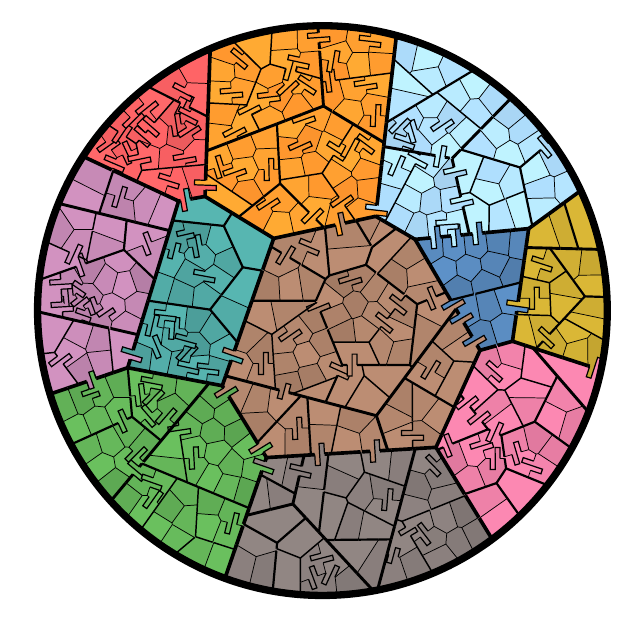}
     \caption{}
         \label{fig:sp500}
         \end{subfigure}
         \begin{subfigure}[b]{0.35\linewidth}
    \centering
    \includegraphics[width=\linewidth]{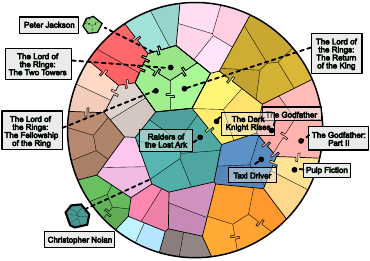}
     \caption{}
         \label{fig:imdb}
\end{subfigure}
\caption{(a) Result of the Flare file hierarchy dataset, with file size encoded as cell weight. The degree of similarity (in this case, code contribution by author) is represented by varying the depth of the interlocking piece. (b) Result of the SP500 dataset showing 429 stocks from 2020 to 2025. The stocks are grouped in Sector, Industry Group, Industry, and Sub-Industry. We calculate the similarities based on the correlation of stock price development.(c) Result of the IMDB dataset. The top level of the Voronoi treemaps shows directors. The individual cells represent movies. We use a binary similarity of the movie genres to calculate the neighborhoods. We see that Peter Jackson's movies share all the same genres, but not Christopher Nolan's.}
\vspace{-3mm}
\end{figure*}
\paragraph*{Country Population (Shared Border)}
This dataset is combined from two sources. The first dataset \cite{datset_countries_population} contains the population of a country, the continent it is located on, and other metadata. The second dataset \cite{datset_countries_borders} contains the borders they share. To make the example more clear, we use only the subset of countries with a population of over 50 million people. Two countries are seen as similar if they share a border. In~\autoref{fig:teaser}, a neighborhood preserving Voronoi treemap is given based on this dataset. Most countries share only borders with other countries on the same continent. But, as shown in the figure, \textit{China} and \textit{Russia} are located on different continents but share a border. The right image of \autoref{fig:algorithm_example} indicates constraints could not be fulfilled with dashed red links.
\paragraph*{Cifar-10 (Language Embeddings)}
A dataset~\cite{krizhevsky2009learning} of images annotated with 10 different class labels, aggregated into a 2-level hierarchy.
Each tree node is enriched with a language embedding using ConceptNet embeddings~\cite{speer2017conceptnet}, which are then aggregated and combined across the hierarchy levels.
The similarity between \textit{bird} and \textit{airplane} makes sense, as they both operate in the sky.

\paragraph*{IMDB Top100 Movies}
We use a subset of 59 of the 100 top-rated movies in the  IMDB dataset. This subset includes only movies of directors who appear at least twice in the IMDB 100 list. \autoref{fig:imdb} shows on the top level the directors and as child cells the individual movies. As the similarity measure, we use a binary attribute, if the genres of two movies are identical. For this dataset, the neighborhood preserving Voronoi treemap shows all movies by Peter Jackson - The Lord of the Rings Trilogy - (green cells in the middle) are all very similar. The movies of other directors like Stanley Kubrick (gray cells to the left) are less similar as they have different genres.
\paragraph*{Flare Data Repository}

The Flare File Hierarchy~\cite{flare-dataset} describes the hierarchical structure of the Flare Repository. The similarity feature on the leaf nodes is the code contribution of different authors, and the size of the nodes is the respective File size. 
As can be observed in~\autoref{fig:flare}, some parts of the hierarchy were created with highly similar co-authors, while others had a lot more variation.

\paragraph*{SP500 Variance}
The SP500 dataset, shown in \autoref{fig:sp500}, contains the development of 429 stocks listed between 2020 and 2025. We used the Global Industry Classification Standard (GICS) to group them hierarchically in \textit{Sector}, \textit{Industry Group}, \textit{Industry}, and \textit{Sub Industry} in a tree. As the similarity measure, we use the correlation of the price development of the individual stocks, which are the individual cells of the Voronoi treemap on the lowest level and the tree leaves.

\section{Discussion and Limitations}\label{sec:discussion}
In this section, we discuss aspects of the similarity graph and visual encoding and highlight our approach's current limitations.

\subsection{Visual Encoding on Edges}

Currently, we encode the similarity of neighboring nodes directly on the edges of the Voronoi cells via the interlocking metaphor. 
However, a node may have more constraints that could not be successfully realized as cell neighbors. 
We show this information on demand by allowing the user to hover over a cell.
The constraints on each node are connected via links.
However, it would also be interesting to see if we can add direct encodings on the cells that highlight how far away these unrealized constraints are. 

We encode the strength of the similarity by three different sizes of our \enquote{Interlocking} metaphor. It might also be an interesting direction to encode different kinds of similarities visually, e.g., to include borders of countries through oceans in \autoref{fig:teaser}.
Further investigation is needed to evaluate different methods of encoding the similarity on edges and how the user perceives this.

\subsection{Similarity Quantification}
As described in~\autoref{sec:preprocessing}, we aggregate, bin, and filter the similarities between nodes.
If we do not filter the similarities between nodes, the similarity graph will oftentimes be fully connected. 
Although we initially tried to work with this fully connected graph during optimization towards their most similar cells, we found that it makes convergence difficult, especially for real-world datasets.  
Therefore, we chose to filter out edges of the similarity graph via binning. 
This, of course, means that we cannot perfectly represent the similarities found in the original data. We argue that by using a relatively lenient automatic binning threshold, we keep a lot of neighborhood relations in not fully connected similarity structures.
However, if the user knows the distribution of the similarities beforehand, it might make sense to allow for custom binning and filtering settings.

\subsection{Planar Embedding considering Similarity}
As noted by Buchin et al.~\cite{similarity_treemaps}, instead of directly creating the Voronoi diagram, we can also construct it indirectly via its dual, the Delaunay graph.
Edges of the Delauney graph become cell edges in the Voronoi diagram.
Therefore, to ensure similar data points share edges in the Voronoi diagram, we could create a graph with edges between similar nodes. 
This graph would have to be planar and triangulated.
Another research area highly connected to this problem is weighted contact representations, where node regions are adjusted to a given node weight~\cite{Noellenburg2013EdgeWeightedContact}. However, node weights are not able to represent individual similarity links, as opposed to methods utilizing edge weights.
We investigated whether or not the problem of finding a suitable similarity graph embedding is connected to the TMFG (Triangulated Maximally Filtered Graph)~\cite{Massara2016NetworkFilteringBig}. 
While the MPS (Maximum Planar Subgraph Problem) is NP-hard in the general case~\cite{michael1979garey}, maximal solutions can be found in polynomial time. 
The TMFG algorithm finds a planar embedding of a weighted edge graph using an iterative clique-based insertion method. However, our case is more complex, as the outerplanar nodes must be within by the parent Voronoi outline. 
As most planar embedding methods, TMFG included, use triangle subdivision to find a suitable embedding, the outerplanar nodes form a triangle, not a polygon. For our use case, all sibling nodes must be within a connected component, and only outer nodes of each component may have an edge connecting them to another parent component. At the same time, the planar embedding of the similarity graph would have to contain all nodes on a hierarchy level. 
We tried utilizing the TMFG approach in our layout step, but found that due to using triangle subdivisions, it cannot be easily adapted to the previously mentioned needed constraints. However, finding a greedy method that produced a planar, filtered similarity graph with the given data similarity and geometric constraints of neighborhood-preserving Voronoi treemaps is still an interesting problem that we might consider in the future. 
\section{Conclusion}
In this paper, we propose neighborhood-preserving Voronoi treemaps, a novel approach to creating treemaps that utilize existing data similarities between nodes to optimize the placement and movement of similar Voronoi cells during optimization so that they share cell edges. 
We found it essential to not only use the similarity to initialize the geometry of the Voronoi diagrams but to actively move cells towards other similar cells during the optimization step if they share a similarity constraint. 
First, data similarities are aggregated and binned to reduce the complexity, creating similarity constraints.
We explored different strategies for the geometric initialization of cells, but found that matching CVT cells with data points based on node similarity and swapping cells to improve neighborhoods gave the best results.
We use similarity constraints to attract nodes during optimization.
This way, we can significantly improve the overall neighborhood preservation of the final result. 
We showcased results from different domains and different complexities. 
To visually convey the similarity to the user, we propose an 'Interlocking' metaphor on the edges of the Voronoi cells.
On hover, all cells linked via constraints are highlighted.
This way, the visualization can be used statically and interactively.
In the future, we hope to evaluate possible strategies to investigate the encoding of similarity on edges further, as well as to preserve more constraints in complex datasets.
Additionally, reducing the algorithmic complexity would greatly improve the applicability of the proposed method to larger real-world datasets.

%% if specified like this the section will be omitted in review mode
\acknowledgments{%
This work was funded by the German Research Foundation (DFG) Project-ID 251654672 – TRR 161 "Quantitative methods for visual computing". Yunhai Wang was supported by the grants of NSFC (No.62132017 and No.U2436209), the Shandong Provincial Natural Science Foundation (No.ZQ2022JQ32), the Fundamental Research Funds for the Central Universities, and the Research Funds of Renmin University of China. 
}

\bibliographystyle{abbrv-doi-hyperref}

\bibliography{template}

\end{document}